\documentclass{ws-ijmpd}

\begin{document}

\markboth{Linqing Wen}
{Detecting Gravitational Waves Using Detector Arrays}

%
\catchline{}{}{}{}{}
%

\title{DATA ANALYSIS OF GRAVITATIONAL WAVES USING A NETWORK OF DETECTORS}

\author{LINQING WEN}

\address{Max Planck Institut f\"ur Gravitationsphysik\\ Albert-Einstein-Institut\\
Am M\"uhlenberg 1\\  D-14476 Golm, Germany\\lwen@aei.mpg.de}%

\maketitle

\begin{history}
\received{14 February 2007}  \comby{W.-T. Ni}
\end{history}

\begin{abstract}
Several large-scale gravitational wave (GW) interferometers have achieved long term operation at design sensitivity.  Questions arise on how to best combine all available data from detectors of different sensitivities for detection, consistency check or veto, localization and waveform extraction.  We show that these problems can be formulated using the singular value decomposition (SVD)\cite{svd} method.  We present techniques based on the SVD method for (1) detection statistic, (2) stable solutions to waveforms, (3) null-stream construction for an arbitrary number of detectors, and (4) source localization for GWs of unknown waveforms.
\end{abstract}

\maketitle

\section{\label{intro} Introduction}

Several large-scale gravitational wave (GW) interferometers have
achieved long term operation at design sensitivity.
Different detectors can have very different noise levels and different
frequency bandwidth.  The directional sensitivity of different
detectors can also be very different.  For instance, the most
sensitive GW detectors in the US, namely the LIGO detector at
Livingston, Louisiana, and the two co-located detectors at Hanford,
Washington (abbreviated as L1 and H1/H2), are designed to be nearly
aligned.  Detectors in Europe (GEO600 in Germany, and VIRGO in Italy)
and in Asia (TAMA in Japan) are incidentally nearly orthogonal in
directional sensitivities to the LIGO detectors.  The questions arise
on how to best combine data from these detectors in GW data analysis.

We present in this paper an application of the singular value
decomposition (SVD) method\cite{svd} to data analysis for a network
of GW detectors. We show that the SVD method provides simple
solutions to detection, waveform extraction, source localization,
and signal-based vetoing. By means of SVD, the response matrix of
the detector network can be decomposed into a product of two unitary
matrices and a pseudo-diagonal matrix containing singular values.
The unitary matrices can be used to form linear combinations of data
from all detectors that have one to one correspondence to linear
combinations of the gravitational wave signal polarization
components. Each newly formed data stream has a corresponding
singular value representing the network's response to the new signal
polarizations. Data streams with non-zero singular values represent
the signal components while data with zero singular values (or zero
multiplication factors) represent the null streams.  The null
streams have null response to GWs and can be used for localization
of GW sources\cite{tinto89} and for identifying detector glitches
from that of real gravitational wave events as proposed by Wen and
Schutz (2005)\cite{wen05} for ground-based GW detectors.  The
statistical uncertainty in estimating the GW waveforms from the data
can be shown to be related to the inverse of the singular values.
This can be used to reduce errors in signal extraction   by enabling
``bad'' data with relatively small singular values to be discarded.


\section{Mathematical Framework}
\label{principle}
The observed strain of an impinging GW by an interferometric detector  is a linear combination of two wave polarizations 
\begin{equation}
h_I(t)=f^{+}_Ih_{+}(t)+f^{\times}_Ih_{\times}(t),
\label{h_t}
\end{equation}
where $t$ is time,  $f^{+}$ and $f^{\times}$ are the detector's  response (antenna beam pattern functions) to the plus and cross
polarizations ($h_+$, $h_{\times}$) of a GW wave.  These antenna beam patterns depend on source sky directions,
wave polarization angle, and detector
orientation.  The $I$ subscript is a label for the
$I$th detector, indicating the dependence of the observed quantities on detectors.


Data from a GW detector can be written as the sum of the detector's response and noise, $d_I(t+\tau_{1I}) = h_I(t)+n_I(t+\tau_{1I})$, where, $t=[0,T]$, $T$ is the duration of the data used, $\tau_{1I}$ is the wavefront arrival time at the $I$-th  detector relative to the reference detector (labeled as detector 1).  The arrival time delay can be calculated as $\tau_{1I}=\hat {\mathbf n} \cdot \mathbf{r_{1I}}/c$, where $\mathbf {r}_{1I}$ is location vector of the $I$-th detector relative to the reference detector, $c$ is the speed of light, ${\hat {\mathbf n}}$ is the unit vector along the wave propagation direction.


For a given source direction, the time-delay corrected data from a network of $N_d$ GW detectors can be written in the frequency domain as,
\begin{equation}
\mathbf{d_k} = A_k\mathbf{h_k}+\mathbf{n_k}, \ \ \  (k=1,..., N_k)
\end{equation}
where, $k$ is the label for frequency bins, $A_k$ is the response matrix of the detector network at each frequency,
\begin{equation}
\mathbf{d_k}=\left ( \begin{matrix}  d_{1k}/\sigma_{1k}\\d_{2k}/\sigma_{2k}\\ ...\\ d_{N_dk}/\sigma_{N_dk}\end{matrix}
\right ),
{\mathbf h_k} =\left ( \begin{matrix} h_{+k}\\ h_{\times k}\end{matrix} \right ),
\mathbf{n_k}=\left ( \begin{matrix}  n_{1k}/\sigma_{1k}\\n_{2k}/\sigma_{2k}\\ ...\\ n_{N_dk}/\sigma_{N_dk}\end{matrix}
\right ),
\end{equation}
and
\begin{equation}
A_k = \left ( \begin{matrix} f^+_1/\sigma_{1k}  & f^\times_1 /\sigma_{1k}  \\ f^+_2/\sigma_{2k}  & f^\times_2/\sigma_{2k} \\ ...& ...\\ f^+_{N_d}/\sigma_{N_d k}  & f^\times_{N_d} /\sigma_{N_d k} \end{matrix}
\right ),
\end{equation}
\label{A_0}
where $\sigma^2_{ik}$ is the noise variance of the $i$-th detector at the $k$-th frequency bin.  The response of the detector network to GWs in all frequencies can be further written in terms of vectors and matrices as,
\begin{equation}
\mathbf{d} = A \mathbf{h}+\mathbf{n},
\end{equation}
where,
\begin{equation}
\mathbf{d}=\left (  \begin{matrix} \mathbf{d_1} \\ \mathbf{d_{2}}\\ ...\\ ...\\ \mathbf{d_{N_k}} \end{matrix}
\right ),
\mathbf{h}=\left (  \begin{matrix} \mathbf{h_1} \\ \mathbf{h_{2}}\\ ...\\ ...\\ \mathbf{h_{N_k}} \end{matrix}
\right ),
\mathbf{n}=\left (  \begin{matrix} \mathbf{n_1} \\ \mathbf{n_{2}}\\ ...\\ ...\\ \mathbf{n_{N_k}} \end{matrix}
\right ),
  A =\left (
\begin{matrix}
 A_1 & & & & \\  & A_2 & & & \\ & & &... &  \\ & & & & A_{N_k}
 \end{matrix}
\right ).
\end{equation}
$A$ of dimension $N_dN_k \times 2N_k$ is therefore the response matrix of the detector network for all frequencies.

\section{Principle of Data Analysis Based on SVD}

The singular value decomposition of the network response matrix $A$ can be written as,
\begin{equation}
A=usv^\dagger, \ \ \ \ s=\left ( \begin{matrix} s_{1} &  0 & ... & 0\\ 0 & s_{2} & ...& 0 \\ 0 &
  ... & ... & ...\\ 0&... &... &s_{2 N_k} \\ 0 & 0 & ... & 0 \\ ... & ... & ... & ... \\ 0 & 0 & ... & 0 \end{matrix} \right ),
\label{svd}
\end{equation}
where singular values $s_{1} \ge s_{2} \ge...\ge s_{2N_k} \ge 0$. $u$ and $v$ are unitary matrices of dimensions $N_dN_k \times N_dN_k$ and $2N_k \times 2N_k$ respectively with $u^\dagger u=I, v^\dagger v=I$.  Note that $s^2_i$ are eigenvalues of $A^{T} A$.  Note also that the singular values of $A$ are only a rearrangement of singular values of $A_k$. 

We propose to construct new data streams by linearly recombining data from different detectors and construct new signal polarizations  using the unitary matrices resulting from the SVD of the network response matrix $A$,
\begin{equation}
\mathbf{d}'=u^\dagger\mathbf{d}, \ \ \ \  \mathbf{h}'=v^\dagger \mathbf{h}, \ \ \ \  \mathbf{n}'=u^\dagger\mathbf{n}.
\end{equation}
These newly constructed data streams and new signal polarizations are one-to-one related to each other via the pseudo-diagonal matrix containing singular values, 
\begin{equation}
\left (  \begin{matrix} {d'_{1}} \\  {d'_{2}}\\ ...\\...\\...\\... \\ {d'_{N_kN_d}} \end{matrix}
\right ) =\left ( \begin{matrix} s_{1} &  0 & ...&0\\ 0 & s_{2} & ...&0 \\ 0&... &... &0\\ 0 &
  ... &...&s_{2N_k}\\ 0 & ... &0 &0\\ ... & ... & ... & ...\\ 0 & ... & ...& 0\\\end{matrix} \right )\left (  \begin{matrix} {h'_{1}} \\  {h'_{2}}\\ ...\\h'_{2N_k}\end{matrix}
\right ) +\left (  \begin{matrix} {n'_{1}} \\  {n'_{2}}\\ ...\\...\\...\\... \\ {n'_{N_kN_d}} \end{matrix}
\right ).
\label{new_data}
\end{equation}
At the presence of a GW signal, there are at most two data streams at each frequency corresponding to non-zero singular values. If the data contain pure stationary Gaussian noise only, the new data stream also follows Gaussian distribution and are statistically uncorrelated with each other.   These data streams can therefore be treated effectively  as one-detector data streams containing signals.  All detection algorithms for a single detector can be applied to these ``signal'' data streams.


The singular values of the response matrix $A$ of the detector network are directly related to the statistical uncertainties in estimating waveforms from the data.  The Fisher information matrix for estimating wave polarizations $\mathbf{h^{'}}$ is,
\begin{equation}
 {I} (\mathbf{h^{'}})=\left ( \begin{matrix} s^2_1 &  0 & ... & 0\\ 0 & s^2_2 & ...& 0 \\ ... &
  ... & ... & ...\\ 0&... &... &s^2_{2N_k}    \end{matrix} \right ),
\label{Fisher}
\end{equation}
where we define the variance of a complex number to be the sum of the variance of the real and complex parts. The lower-bound\cite{cramer46} for the variance of the estimated GW signals $\mathbf{h}_k$ due to statistical errors are therefore a linear combination of $s^{-2}_i$.  Singular values are therefore indicators for the accuracy of waveform estimation and can be used for regularization of the solutions to the waveforms.

At the end, data streams corresponding to zero singular values or zero multiplication factors have null response to signals (Eq.~\ref{new_data}). There are at least $(N_d-2)N_k$ ``null streams'' due to the fact that there are only two polarizations for a plane GW based on Einstein's theory of General Relativity.  Null streams can therefore be used to test the consistency of the detected GW events  and to veto against signal-like noise.\cite{wen05}

\subsection{Detection Statistic}
\label{detection}
In this section, we discuss the construction of detection statistic for GWs of unknown waveforms using data from a network of detectors.   We discuss two possible approaches,  (1) a direct application of  maximum likelihood ratio\cite{mlr} (MLR) principle by constructing detection statistic from adding powers of all data with non-zero singular values, (2) optimized detection statistic based on MLR together with assumptions on wave spectrum and principle of maximum signal-to-noise ratio (SNR). SNR is a good indicator for detection efficiency in our case as, when adding sufficient large number of noise powers together, central limit theorem states that these SNRs, regardless of underlying noise distributions, follow roughly Gaussian distribution.

The ``standard'' solution is based on a direct application of the MLR principle for unknown waveforms.  When the signal polarizations $h_{+k}$ and $h_{\times k}$ are assumed to be independent variables for all frequencies, the components of the new signal $\mathbf{h}'$ are also independent from one another.   The standard MLR detection statistic can then be constructed by adding equally all powers of ``signal'' data-streams of non-zero singular values,
\begin{equation}
P^{(0)}_S=\sum^{N_p}_{i=1} |d'_i|^2,
\label{P0_det}
\end{equation}
where $N_p$ is the number of non-zero singular values. Note that singular values have already been ranked (Eq.~\ref{svd}).  In the absence of noise, $|d'_i|^2=s^2_i |h'|^2_i$.   In the presence of pure stationary Gaussian noise, $2|d'_i|^2$ follows $\chi^2_2$ distribution with variance of $4$.   By discarding data of null response to GW signals, this detection statistic is optimal in SNRs than simply adding powers of all (noise-weighted) data together.   However, in the presence of extremely small singular values (e.g, when two detectors are nearly aligned) or extremely weak signals, this detection statistic does not necessarily maximize the SNR.

Singular values can be used to further optimize the detection statistic by making reasonable assumptions about the signal spectrum. For instance, we can omit signal powers of small singular values with hopes that the chance is very small for the new signal component $\mathbf{h'}$  to be very strong at frequencies and polarizations where the network is not sensitive.  Under this assumption, the detection statistic can be,
\begin{equation}
P^{(1)}_S=\sum^{N'_p}_{i=1} |\mathbf{d'}_{i}|^2.
\label{p1_det}
\end{equation}
That is, only $N'_p \le N_p $ number of powers of reasonable large singular values are included. 

If we assume that the detection statistic is a linear combination of individual powers and that the signal power $|h^{'}_i|^2$ is the same in both frequency and new polarizations, maximization of the SNR leads to,
\begin{equation}
P^{(2)}_S=\sum^{N'_p}_{i=1} \alpha_i |(\mathbf{d'})_{i}|^2,
\label{p2_det}
\end{equation}
where,
\begin{equation}
\alpha_{i }=\frac{s^2_{i}}{\sqrt{\sum_{j} s^4_{i}}}.
\label{p1_alpha}
\end{equation}
If we assume flat power in signal polarizations only, summation and normalization of $s^2_i$ terms in Eq.~\ref{p1_alpha} apply only to the two polarizations within each frequency.


We have also investigated the approach of adding powers within a window of the ranked singular values as a general approach to unknown waveforms,
\begin{equation}
P^{(3)}_S=\sum^{i_1+N'}_{i=i_1} |d'_{i}|^2.
\label{p3_det}
\end{equation}
where the bandwidth of $N'$ is determined by maximizing
the SNR,
\begin{equation}
SNR_{N^{'}} =\mbox{max}_{N^{'}} (P_{N^{'}} /\sqrt{N^{'}}).
\end{equation}
This is an ad hoc approach that works better for signal powers that are narrow-banded in frequencies.
While this approach can be time-consuming, it performs pretty well in our test with the BH-BH merger signals.

Early work on optimization of the detection statistic based on directional sensitivity of GW detectors can be found in Rajesh Nayak et al. (2003) for future space detector LISA\cite{pai03} and  in Klimenko et al. (2005, 2006) for ground-based detectors.\cite{klimenko05,klimenko06}  It has been demonstrated in these papers that further optimization from the standard MLR detection statistic can be obtained by constraining or eliminating contributions of data corresponding to weaker network sensitivity.   
Other work can be found in Mohanty et al. (2006)\cite{mohanty06} which is based on the SNR variability regulator and in Summerscales et al. (2006)\cite{finn06} which is based on the maximum entropy principle.


We show that the SVD method provides a simple mean to the construction of the detection statistic and its further optimization.  The standard detection statistic based on the maximum likelihood ratio principle can be easily reproduced from the newly constructed data.  Various optimization strategies can then be obtained based on general assumptions of the signals together with the singular values which represent the sensitivity of the detector network.  We have also proposed a strategy by maximization of the SNRs from the summation of powers corresponding to a window of ranked singular values.     A comparison of the performance of different detection statistic can be found in a follow-up paper.\cite{wen07a}

\subsection{Stable Solutions to Waveform Extraction}
The ``standard'' estimator for $\mathbf{h'}=[h'_1,...,h'_{N_p}]^T$ according to the maximum likelihood ratio principle of ``burst'' GWs (defined as any GWs of unknown waveforms) is
\begin{equation}
{h^{'}}_i=s^{-1}_i {d^{'}}_i, \ \ \ \ i=1,..., N_p,
\label{h_k1}
\end{equation}
where $N_p$ is the number of non-zero singular values. Note the components of $h'$ include two polarization components for each frequencies.  The ``standard'' solutions to wave polarizations can be extracted using the SVD components of the network response (Eq.~\ref{svd}, Eq.~\ref{new_data}) 
\begin{equation}
\mathbf{h}^{(0)}=v \mathbf {h'}.
\label{h_k0}
\end{equation}
Note that the solutions of two signal polarization components  at each frequency depends on quantities within that frequency only.  Faster calculations can therefore be carried out independently at each frequencies. In the limit when $N_p =2N_k$, the solution given in Eq.~\ref{h_k0} is equivalent to that written with the Moore-Penrose inverse,
$\mathbf{h}^{(0)}= (A^\dagger A)^{-1}A^\dagger \mathbf{d}$.
Note that this is also the same as the effective one-detector strain
for data from a network of GW detectors introduced by Flanagan and
Hughes (1998).\cite{flanagan98}

The ``standard'' solution $\mathbf {h}^{(0)}$ from Eq.~\ref{h_k0} however can possibly lead to unstable solutions in the sense that a small error in the data can lead to large amplified error in the solution.  Fisher information matrix (Eq.~\ref{Fisher}) indicates that the best possible statistical variance for the estimated $h^0_k$ is a linear combination of $1/s^2_i$.
The extracted  wave polarizations $h^0_k$ can contain large errors if we include data corresponding to very small singular values. This situation occurs when the response matrix $A$ is ``ill-conditioned'' with $s_{i}/s_1 \ll 1$.  The small singular values can result from machine truncation errors instead of zero values or  from nearly degenerated solutions to the equations, e.g., in our case, when antenna beam patterns of two detectors are nearly aligned.

Regularization is needed in order to have stable solutions to
$\mathbf{h}_k$.   One simple solution is to apply the ``truncated singular
value decomposition'' (TSVD) method by omitting data with very small singular values in Eq.~\ref{h_k0}.
\begin{equation}
{h}^T_i=\sum^{N_T}_{j=1}v_{ij} h'_{j},
\label{h_T}
\end{equation}
where $N_T \le N_p$ is the number of data included. The main problem is then the decision on where to start to truncate the data streams with small singular
values. It depends on the accuracy requirement in waveform extraction, type of waveforms and type of constraints that can be put on the solutions.  The
fractional error due to truncation, defined as the ratio of the sum of error-square
and the sum of detector-response square from all frequencies, is,
\begin{equation}
\frac {||Ah-A\mathbf {h^{T}}||}{||A\mathbf {h}||}
=\frac{\sum^{N_p}_{j=N_T+1} s^2_{j}|(v^\dagger\mathbf h)_{j}|^2}{\sum^{N_p}_{j=1}s^2_{j}| (v^\dagger\mathbf h)_{j}|^2}
\label{LS}
\end{equation}
Truncation of terms with very small singular values can therefore retain the least square fit of the detector response to the data while greatly reduce the statistical errors when extracting individual signal polarizations.  A recent work on introducing the Tikhonov regularization to waveform extraction of GW signals can be found in Rakhmanov (2006).\cite{malik06}  Note that the Tikhonov regularization is equivalent to introducing a new parameter to filtering out data associated with small singular values.

\subsection{Null Stream Construction}
\label{null}
There are at least $(N_d-2)N_k$ data streams with zero singular
values or zero multiplication factors. These are data streams that
have null response to signals.  The ``standard'' null streams can be
written in terms of the SVD of $A$ (Eq.~\ref{svd}) as,
\begin{equation}
N^0_i=(u^\dagger \mathbf{d})_{i+N_p}, \ \ \ \ i=1, 2, ..., N_dN_k-N_p.
\label{null_data}
\end{equation}
For stationary Gaussian noise, $N^0_i$ follows Gaussian distribution
of zero mean and unity variance and are statistically independent
with each other.   Consistency check for a detected GW event
therefore requires that all the null-streams are consistent with
`noise-only'' at the source direction.  Using null-streams as a tool
for consistency check of GW events against signal-like glitches for
ground-based GW detectors was first proposed by Wen and Schutz
(2005).\cite{wen05}  Further investigations on consistency check
using the null-streams and its relation to the null space of the
network-response matrix $A$ can be found in Chatterji et al.
(2006).\cite{caltech06}


There are also ``semi-'' null streams corresponding to data streams with very small but non-zero singular values,
\begin{equation}
N_{Wi}=(u^\dagger \mathbf{d})_{i+N'_p-1}, \ \ \ \ i=1, 2, ..., N_p-N'_p,
\label{semi_null}
\end{equation}
where $N'_p$ is the starting indexes for $1\gg s_i >0$. These semi-null streams  exist when the equations are nearly degenerated.  This happens, for example, for response of the two LIGO detectors of H1 (H2) and L1 that are designed to be nearly aligned.  The semi-null stream can be also simply caused by combination of weak directional sensitivity and/or high noise level instrument of all detectors in the network.   Consistency check  of GW events, veto against noise,  and localization of the source can be further improved by including both the ``standard'' null-streams of analytically zero singular values and also these semi-null streams.
\begin{equation}
P^{1}_{N}=\sum^{N_dN_k-Np}_{i=1} |N^0_i|^2+\sum^{N_p-N'_p}_{i=1} |{N_W}_i|^2.
\end{equation}
The usage of semi-null streams depends on waveforms and therefore
efficiency studies should be carried out before-hand.  A proposal of
using the semi-null stream for veto against detector glitches and
for source localization can be found in Wen and Schutz
(2005)\cite{wen05} for the two-detector network of H1-L1.

\section{Conclusion}
\label{conclusion} A new data analysis approach based on the
singular value decomposition method has been proposed for the data
analysis of GWs using a network of detectors.  We show that singular
values of the response matrix of the GW network directly encode the
sensitivity of the GW network to signals and the uncertainties in
waveform estimation.  The SVD method is particularly useful for
constructing null data streams that have no or very little response
to GW signals.  We argue that the SVD method provide a simple recipe
for data analysis of GWs for a network of detectors.  Note that the
SVD method is widely used in engineering signal processing for image
processing, noise reduction and geophysical inversion problems. An
application of the SVD method to detecting GWs from the
extreme-mass-ratio-inspiral sources using the space GW detector LISA
can be found in Wen et al. (2006).\cite{wen06} The SVD software
package is also available in MATLAB.

Strategies on construction of the detection statistic, stable solutions to waveforms, and null streams based on the SVD method are discussed.   We show that the detection statistic should be constructed from data streams of non-zero singular values. We discuss how detection efficiency can be improved from a direct application of MLR by incorporating our knowledge of the network's directional response to GWs and our assumptions on distribution of the signal power.

We also give expressions of null-streams for arbitrary number of GW
detectors using components from the SVD of the network-response
matrix.  The concept of the semi-null streams that are characterized
by small singular values is also introduced. We argue that the
exploration of the null-space by including semi-null streams can
help improving the source localization and consistency check.
Analytical study for the angular resolution of a network of GW
detectors and their relations with the null space will be found in
Wen\cite{wen07b} (2007). We also show how a stable solution to the
waveform can be constructed based on information from the singular
values.

We conclude that a GW event should be identified only when both of the following two conditions are satisfied at the same sky direction, (1) high probability of detection for the optimal statistic constructed from ``signal'' streams with non-zero singular values (Eq.~\ref{new_data}, sec.~\ref{detection}) and (2) high probability that the null streams corresponding to zero singular values are consistent with noise (Eq.~\ref{new_data}, Eq.~\ref{null_data}, sec.~\ref{null}).
Results of an extended investigation will be published elsewhere.\cite{wen07a}

\section*{Acknowledgments}

We are very grateful to Yanbei Chen for his critical discussions. We also thank Wei-Tou Ni, Sergey Klimenko, Soumya Mohanty, and David Blair for careful review and helpful comments of this manuscript.  This work is under the support by the Alexander von Humboldt Foundation's Sofja Kovalevskaja Programme funded by the German Federal Ministry of Education and Research. This paper has been assigned LIGO document number LIGO-P070012-00-Z.


\end{document}